\documentclass[a4paper,12pt,fleqn]{article} 

\usepackage{epsfig}
\usepackage{amsmath}
\usepackage{amssymb}
\usepackage{mathrsfs}
\usepackage{amsfonts}

\begin{document}

{
\center 

{\Large Signal discovery in sparse spectra: a Bayesian analysis}

\vspace{1.0cm} 

A.~Caldwell, K.~Kr\"oninger 

\vspace{0.5cm} 

{\it Max-Planck-Institut f\"ur Physik, M\"unchen, Germany} 

\vspace{0.5cm}

25.08.2006

} 

\begin{abstract} 
A Bayesian analysis of the probability of a signal in the presence of
background is developed, and criteria are proposed for claiming
evidence for, or the discovery of a signal.  The method is general and
in particular applicable to sparsely populated spectra. Monte Carlo
techniques to evaluate the sensitivity of an experiment are
described. \\

As an example, the method is used to calculate the sensitivity of the
{\sc GERDA} experiment to neutrinoless double beta decay.
\end{abstract} 


\section{Introduction}
\label{section:introduction}

In the analysis of sparsely populated spectra common approximations,
valid only for large numbers, fail for the small number of events
encountered. A Bayesian analysis of the probability of a signal in the
presence of background is developed, and criteria are proposed for
claiming evidence for, or the discovery of a signal.  It is
independent of the physics case and can be applied to a variety of
situations. 

To make predictions about possible outcomes of an experiment,
distributions of quantities under study are calculated. As an
approximation, ensembles, sets of Monte Carlo data which mimic the
expected spectrum, are randomly generated and analyzed. The frequency
distributions of output parameters of the Bayesian analysis are
interpreted as probability densities and are used to evaluate the
sensitivity of the experiment to the process under study.  

As an example, the analysis method is used to estimate the sensitivity
of the {\sc GERDA} experiment~\cite{proposal} to neutrinoless double
beta decay. 

The analysis strategy is introduced in section~\ref{section:analysis}.
The generation of ensembles and the application of the method onto
those is discussed in section~\ref{section:ensembletests}. The
application of the analysis method in the {\sc GERDA} experiment is
given as an example in section~\ref{section:sensitivity} where the
sensitivity of the experiment is evaluated.


\section{Spectral analysis}
\label{section:analysis}
 
A common situation in the analysis of data is the following: two types
of processes (referred to as signal and background in the following)
potentially contribute to a measured spectrum. The basic questions
which are to be answered can be phrased as: {\it What is the
contribution of the signal process to the observed spectrum? What is
the probability that the spectrum is due to background only? Given a
model for the signal and background, what is the (most probable)
parameter value describing the number of signal events in the
spectrum? In case no signal is observed, what is the limit that can be
set on the signal contribution?}  The analysis method introduced in
this paper is based on Bayes' Theorem and developed to answer these
questions and is in particular suitable for spectra with a small
number of events.

\noindent  
The assumptions for the analysis are

\begin{itemize}
\item The spectrum is confined to a certain region of interest. 
\item The spectral shape of a possible signal is known. 
\item The spectral shape of the background is known\footnote{This assumption and the previous can be removed 
in a straightforward way with the introduction of additional prior
densities.}.
\item The spectrum is divided into bins and the event numbers in 
the bins follow Poisson distributions.
\end{itemize} 

The analysis consists of two steps. First, the probability that the
observed spectrum is due to background only is calculated. If this
probability is less then an {\it a priori} defined value, the
discovery (or evidence) criterion, the signal process is assumed to
contribute to the spectrum and a discovery (or evidence) is
claimed. If the process is known to exist, this step is skipped. Based
on the outcome, in a second step the signal contribution is either
estimated or an upper limit for the signal contribution is calculated.

\subsection{Hypothesis test}

In the following, $H$ denotes the hypothesis that the observed
spectrum is due to background only; the negation, interpreted here as
the hypothesis that the signal process contributes to the
spectrum\footnote{Since the shape of the background spectrum is
assumed to be known the case of unknown background sources
contributing to the measured spectrum is ignored. However, the overall
level of background is allowed to vary.}, is labeled
$\overline{H}$. The conditional probabilities for the hypotheses $H$
and $\overline{H}$ to be true or not, given the measured spectrum are
labeled $p(H|\textrm{spectrum})$ and
$p(\overline{H}|\textrm{spectrum})$, respectively. They obey the
following relation:
\begin{equation}
p(H|\textrm{spectrum}) + p(\overline{H}|\textrm{spectrum}) = 1\ . 
\end{equation}

The conditional probabilities for $H$ and $\overline{H}$ can be
calculated using Bayes' Theorem~\cite{Bayes}:
\begin{equation}
p(H|\textrm{spectrum}) = \frac{p(\textrm{spectrum}|H) \cdot p_{0}(H|I)}{p(\mathrm{spectrum})} 
\label{eqn:pH}
\end{equation} 

and 
\begin{equation}
p(\overline{H}|\textrm{spectrum}) = \frac{p(\textrm{spectrum}|\overline{H}) \cdot p_{0}(\overline{H}|I)}{p(\mathrm{spectrum})}, 
\label{eqn:pHbar}
\end{equation} 

\noindent 
where $p(\textrm{spectrum}|H)$ and $p(\textrm{spectrum}|\overline{H})$
are the conditional probabilities to find the observed spectrum given
that the hypothesis $H$ is true or not true, respectively and
$p_{0}(H|I)$ and $p_{0}(\overline{H}|I)$ are the prior probabilities
for $H$ and $\overline{H}$. The values of $p_{0}(H|I)$ and
$p_{0}(\overline{H}|I)$ are chosen depending on additional
information, $I$, such as existing knowledge from previous experiments
and model predictions. In the following, the symbol $I$ is dropped but
it should be understood that all available information is used in the
evaluation of probabilities. The probability $p(\mathrm{spectrum})$ is
rewritten as
\begin{equation}
p(\textrm{spectrum}) = p(\textrm{spectrum}|H) \cdot p_{0}(H) + p(\textrm{spectrum}|\overline{H}) \cdot p_{0}(\overline{H}) 
\end{equation} 

The probabilities $p(\textrm{spectrum}|H)$ and
$p(\textrm{spectrum}|\overline{H})$ can be decomposed in terms of the
expected number of signal events, $S$, and the expected number of
background events, $B$:
\begin{eqnarray}
p(\textrm{spectrum}|H)       & = & \int p(\textrm{spectrum}|B) \cdot p_{0}(B)~dB, 
\label{eqn:expansionH} \\ 
p(\textrm{spectrum}|\overline{H}) & = & \int p(\textrm{spectrum}|S,~B) \cdot p_{0}(S) \cdot p_{0}(B)~dS~dB, 
\label{eqn:expansionHbar} 
\end{eqnarray}

\noindent
where $p(\textrm{spectrum}|B)$ and $p(\textrm{spectrum}|S,~B)$ are the
conditional probabilities to obtain the measured spectrum. Further,
$p_{0}(S)$ and $p_{0}(B)$ are the prior probabilities for the number
of signal and background events, respectively. They are assumed to be
uncorrelated, and are chosen depending on the knowledge from previous
experiments, supporting measurements and models.

The {\it observed} number of events in the $i$th bin of the spectrum
is denoted $n_{i}$. Assuming the fluctuations in the bins of the
spectrum to be uncorrelated the probability to observe the measured
spectrum, given $B$ (in case $H$ is true) or the set $S$, $B$ (in case
$\overline{H}$ is true), is simply the product of the probabilities to
observe the $N$ values, $\{n_{i}\}$. The {\it expected} number of
events in the $i$th bin, $\lambda_{i}$, can be expressed in terms of
$S$ and $B$:
\begin{eqnarray}
\label{eqn:lambda}
\lambda_{i} & = & \lambda_{i}(S,~B) \\ 
      & = & S \cdot \int_{\Delta E_{i}} f_{\mathrm{S}}(E)~dE + B \cdot \int_{\Delta E_{i}} f_{\mathrm{B}}(E)~dE, \nonumber 
\end{eqnarray} 

\noindent 
where $f_{\mathrm{S}}(E)$ and $f_{\mathrm{B}}(E)$ are the normalized
shapes of the known signal and background spectra, respectively, and
$\Delta E_{i}$ is the width of the $i$th bin. The letter $E$ suggests
an energy bin, but the binning can be performed in any quantity of
interest. The number of events in each bin can fluctuate around
$\lambda_{i}$ according to a Poisson distribution. This yields
\begin{eqnarray}
\label{eqn:pbn}
 p(\textrm{spectrum}|B)    & = & \prod_{i = 1}^{N} \frac{\lambda_{i}(0,~B)^{n_{i}}}{n_{i}!} e^{-\lambda_{i}(0,~B)} \\ 
\label{eqn:psn}
p(\textrm{spectrum}|S,~B) & = & \prod_{i = 1}^{N} \frac{\lambda_{i}(S,~B)^{n_{i}}}{n_{i}!} e^{-\lambda_{i}(S,~B)}\ . 
\end{eqnarray} 

In summary, the probability for $H$ to be true, given the measured
spectrum, is:
{\small
\mathindent = 0.pt 
\begin{eqnarray}
\lefteqn{p(H|\textrm{spectrum}) =} \nonumber \\
\\ 
\lefteqn{ \frac{\left[\int \prod \frac{\lambda_i^{n_{i}}}{n_{i}!} e^{-\lambda_i} \cdot p_{0}(B)~dB\right]_{S=0} \cdot p_{0}(H)} {\left[ \int \prod  \frac{\lambda_i^{n_{i}}}{n_{i}!} e^{-\lambda_i} \cdot p_{0}(B)~dB\right] _{S=0}\cdot p_{0}(H)+ \left[\int \prod \frac{\lambda_i^{n_{i}}}{n_{i}!} e^{-\lambda_i} \cdot p_{0}(B)p_{0}(S)~dB~dS\right] \cdot p_{0}(\overline{H})}  \nonumber}
\end{eqnarray}
\mathindent = 25.pt
}

\noindent 
with $\lambda_i$ calculated according to (\ref{eqn:lambda}).  Evidence
for a signal or a discovery can be decided based on the resulting
value for $p(H|\textrm{spectrum})$.  It should be emphasized that the
discovery criterion has to be chosen {\it before} the data is
analyzed. A value of $p(H|\textrm{spectrum})\leq 0.0001$ is proposed
for the \emph{discovery} criterion, whereas a value of
$p(H|\textrm{spectrum})\leq 0.01$ can be considered to give
\emph{evidence} for $\overline{H}$.

The analysis can be easily extended to include uncertainties in the
knowledge of relevant quantities.  For example, if the spectrum is
plotted as a function of energy, and the energy scale has an
uncertainty, then equations~(\ref{eqn:pbn},\ref{eqn:psn}) can be
rewritten as
\begin{eqnarray}
 p(\textrm{spectrum}|B)    & = & \int \left[ \prod_{i = 1}^{N} \frac{\lambda_{i}(0,~B|k)^{n_{i}}}{n_{i}!} e^{-\lambda_{i}(0,~B|k)}\right] p_0(k) dk \\ 
p(\textrm{spectrum}|S,~B) & = & \int \left[ \prod_{i = 1}^{N} \frac{\lambda_{i}(S,~B|k)^{n_{i}}}{n_{i}!} e^{-\lambda_{i}(S,~B|k)}\right] p_0(k) dk\ . 
\end{eqnarray} 

\noindent 
where $\lambda_{i}(S,~B|k)$ is the expected number of events for a given energy scale factor $k$ and $p_0(k)$ is the probability density for $k$ (e.g., a Gaussian distribution centered on $k=1$).

\subsection{Signal parameter estimate} 

In case the spectrum fulfills the requirement of evidence or
discovery, the number of signal events can be estimated from the
data. The probability that the observed spectrum can be explained by
the set of parameters $S$ and $B$, making again use of Bayes' Theorem,
is:
\begin{equation}
p(S,\ B|\textrm{spectrum}) = \frac{p(\textrm{spectrum}|S,~B)
\cdot p_{0}(S) \cdot p_{0}(B)}{\int p(\textrm{spectrum}|S,~B) \cdot
p_{0}(S) \cdot p_{0}(B)~dS~dB}\ .
\end{equation} 

In order to estimate the signal contribution the probability
$p(S,~B|\textrm{spectrum})$ is marginalized with respect to $B$:
\begin{equation} 
p(S|\textrm{spectrum}) = \int p(S,~B|\textrm{spectrum})~dB\ . 
\label{eqn:pSSpectrum} 
\end{equation} 

The mode of this distribution, $S^*$, i.e., the value of $S$ which
maximizes $p(S|\textrm{spectrum})$, can be used as an estimator for
the signal contribution. The standard uncertainty on $S$ can be
evaluated from
\begin{eqnarray*}
\int_{0}^{S_{16}} p(S|\textrm{spectrum})~dS & =&  0.16 \\
\int_{0}^{S_{84}} p(S|\textrm{spectrum})~dS & =&  0.84 \\
\end{eqnarray*}
such that the results can be quoted as
{\large
\begin{equation}
{S^*} ^{+(S_{84}-S^*)}_{-(S*-S_{16})} \; .
\end{equation} 
}

\subsection{Setting limits on the signal parameter} 

In case the requirement for an observation of the signal process is
not fulfilled an upper limit on the number of signal events is
calculated. For example, a 90\% probability lower limit is calculated
by integrating Equation~(\ref{eqn:pSSpectrum}) to 90\% probability:
\begin{equation}
\int_{0}^{S_{90}} p(S|\textrm{spectrum})~dS = 0.90 \ . 
\end{equation} 

$S_{90}$ is the 90\% probability upper limit on the number of signal
events. It should be noted that in this case it is assumed that
$\overline{H}$ is true but the signal process is too weak to
significantly contribute to the spectrum.


\section{Making predictions - ensemble tests}
\label{section:ensembletests}

\begin{figure}[ht!]
\epsfig{file=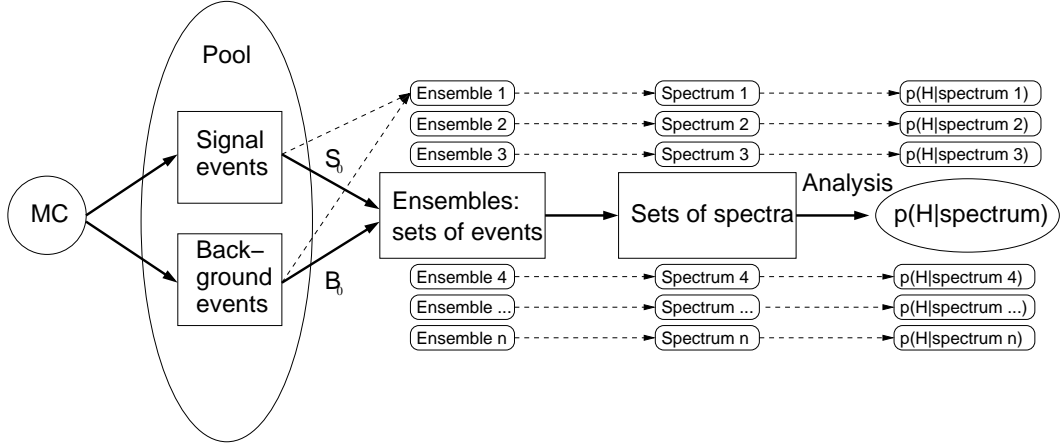,width=\textwidth}
\caption{Analysis chain. The Monte Carlo generator (MC) generates a  
pool which consists of signal and background events. An ensemble is
defined as a set of events representing a possible outcome of an
experiment. The number of events are randomly chosen according to the
parameters $S_{0}$ and $B_{0}$. From each ensemble a spectrum is
extracted and subsequently analyzed. The probability
$p(H|\textrm{spectrum})$ for each spectrum is depicted here as the
outcome of the analysis.
\label{fig:chain}}
\end{figure}

In order to predict the outcome of an experiment distributions of the
quantities under study can be calculated. This is done numerically by
generating possible spectra and subsequently analyzing these. The
spectra are typically generated from Monte Carlo simulations of signal
and background events.  For a given ensemble, the number of signal and
background events, $S_{0}$ and $B_{0}$ are fixed and a random number
of events are collected according to Poisson distributions with means
$S_0$ and $B_0$.  From each ensemble a spectrum is extracted and the
analysis described above is applied. The analysis chain is shown in
Figure~\ref{fig:chain}. 
 
The output parameters, such as the conditional probability for $H$, \linebreak 
$p(H|\textrm{spectrum})$, are histogrammed and the frequency
distribution is interpreted as the probability density for the
parameter under study. As examples, the mean value and the
16\%~to~84\% probability intervals can be deduced and used to predict
the outcome of the experiment. This approach is referred to as
ensemble tests.  
 
Systematic uncertainties, such as the influence of energy resolution,
miscalibration or signal and background efficiencies, can be estimated
by analyzing ensembles which are generated under different
assumptions.

\section{Sensitivity of the {\sc GERDA} experiment}
\label{section:sensitivity}

In the following, the {\sc GERDA} experiment is introduced and the
Bayesian analysis method, developed in section~\ref{section:analysis},
is applied on Monte Carlo data in order to predict possible outcomes
of the experiment.

\subsection{Neutrinoless double beta decay and the {\sc GERDA} experiment} 
 
The GERmanium Detector Array, {\sc GERDA}~\cite{proposal}, is a new
experiment to search for neutrinoless double beta decay
(0$\nu\beta\beta$) of the germanium isotope $^{76}$Ge.  Neutrinoless
double beta decay is a second order weak process which is predicted to
occur if the neutrino is a Majorana particle. The half-life of the
process is a function of the neutrino masses, their mixing angles, and
the CP phases. Today, 90\%~C.L. limits on the half-life for
neutrinoless double beta decay of $^{76}$Ge exist and come from the
Heidelberg-Moscow~\cite{HM} and IGEX~\cite{IGEXlimit}
experiments. They are $T_{1/2}>1.9\cdot10^{25}$~years and
$T_{1/2}>1.6\cdot10^{25}$~years, respectively. A positive claim was
given by parts of the Heidelberg-Moscow collaboration with a
$3~\sigma$ range of $T_{1/2}=(0.7-4.2)\cdot10^{25}$~years and a best
value of $T_{1/2}=1.2\cdot10^{25}$~years~\cite{Klapdor}. \\

A total exposure (measured in kg$\cdot$years of operating the
germanium diodes) of at least 100~kg$\cdot$years should be collected
during the run-time of the {\sc GERDA} experiment. The germanium
diodes are enriched in the isotope $^{76}$Ge to a level of about
86\%. One of the most ambitious goals of the experiment is the
envisioned background level of
$10^{-3}$~counts/(kg$\cdot$keV$\cdot$y). This is two orders of
magnitude below the background index observed in previous
experiments~\cite{HM,IGEX}. For an exposure of 100~kg$\cdot$years the
expected number of background events in the 100 keV wide region of
interest is approximately 10. Using the present best limit on the
half-life less than 20 $0\nu\beta\beta$-events are expected within a
much smaller window. The number of expected $0\nu\beta\beta$-events,
$S_{0}$, is correlated with the half-life of the process via
\begin{equation}
S_{0} \approx \ln2 \cdot \kappa \cdot M \cdot \epsilon_{\mathrm{sig}} \cdot \frac{N_{\mathrm{A}}}{M_{\mathrm{A}}} \cdot \frac{t}{T_{1/2}}, 
\label{eqn:halflife} 
\end{equation}

\noindent 
where $\kappa = 0.86$ is the enrichment factor, $M$ is the mass of
germanium in grams, $N_{\mathrm{A}}$ is Avogadro's constant and $t$ is
the measuring time. $M_{\mathrm{A}}$ is the atomic mass and
$\epsilon_{\mathrm{sig}}$ is the signal efficiency, estimated from
Monte Carlo data to be 87\%.

\subsection{Expected spectral shapes and prior probabilities} 
 
In {\sc GERDA}, the energy spectrum in the region around 2~MeV is
expected to be populated by events from various background
processes. The signature of neutrinoless double beta decay, the signal
process, is a sharp spectral line at the $Q_{\beta\beta}$-value which
for the germanium isotope $^{76}$Ge is $2\,039$~keV. In the following,
the region of interest is defined as an energy window of $\pm 50$~keV
around the $Q_{\beta\beta}$-value. The shape of the background
spectrum is assumed to be flat, i.e. $f_{\mathrm{B}}(E)=const$. The
shape of the signal contribution is assumed to be Gaussian with a mean
value at the $Q_{\beta\beta}$-value. The energy resolution of the
germanium detectors in the {\sc GERDA} setup is expected to be 5~keV
(FWHM), corresponding to a width of the signal Gaussian of $\sigma
\approx 2.1$~keV. 

For the calculation of the sensitivity, ensembles are generated
according to (1) the exposure, (2) the half-life of the
$0\nu\beta\beta$-process which is translated into the number of
expected signal events, $S_{0}$, in the spectrum, and (2) the
background index in the region of interest which is translated into
the number of expected background events, $B_{0}$.  The number of
signal and background events in each ensemble fluctuate around their
expectation values $S_{0}$ and $B_{0}$ according to a Poisson
distribution. For each set of input parameters 1000 ensembles are
generated. An energy spectrum is extracted from each ensemble with a
bin size of 1~keV.

In order to calculate the probability that the spectrum is due to
background processes only, the prior probabilities for the hypothesis
$H$ and $\overline{H}$ have to be fixed, as well as those for the
signal and background contributions. This is a key step in the
Bayesian analysis.  Given the lack of theoretical consensus on the
Majorana nature of neutrinos and the cloudy experimental picture, the
prior probabilities for $H$ and $\overline{H}$ are chosen to be equal,
i.e.
\begin{eqnarray}
p_{0}(H)       & = & 0.5, \\ 
p_{0}(\overline{H}) & = & 0.5\ . 
\end{eqnarray}

The prior probability for the number of expected signal events,
assuming $\overline{H}$, is assumed flat up to a maximum value,
$S_{max}$, consistent with existing limits\footnote{$S_{\mathrm{max}}$
was calculated using Equation~\ref{eqn:halflife} assuming a half-life
of $T_{1/2}=0.5\cdot10^{25}$~years.}.  It should be noted that the
setting of the prior probability for $H$ is dependent on the maximum
allowed signal rate.  $S_{\mathrm max}$ was chosen in such a way that
the probability for the hypothesis $H$ is reasonably assumed to be
50~\%. The effect of choosing a different prior for the number of
signal events is discussed below.

The background contribution $B$ is assumed to be known within some
uncertainty (recall that the shape of the background is however
fixed). The prior probability for $B$ is chosen to be Gaussian with
mean value $\mu_{\mathrm{B}}=B_{0}$ and width
$\sigma_{\mathrm{B}}=B_{0}/2$. The prior probabilities for the
expected signal and background contributions are
\begin{eqnarray}
p_{0}(S) = \frac{1}{S_{\mathrm{max}}},\ 0\le S\le S_{\mathrm{max}},\ p_{0}(S)=0\ \textrm{otherwise}, \\
p_{0}(B) = \frac{e^{-\frac{(B-\mu_{\mathrm{B}})^{2}}{2 \sigma_{\mathrm{B}}^{2}}}}{\int_{0}^{\infty}e^{-\frac{(B-\mu_{\mathrm{B}})^{2}}{2 \sigma_{\mathrm{B}}^{2}}}},\ B\ge 0,\ p_{0}(B)=0\ B<0\ . 
\end{eqnarray} 

\subsection{Examples} 
 
\begin{figure}[ht!]
\begin{tabular}{cc}
\begin{minipage}[ht!]{0.48\textwidth}
\mbox{\epsfig{file=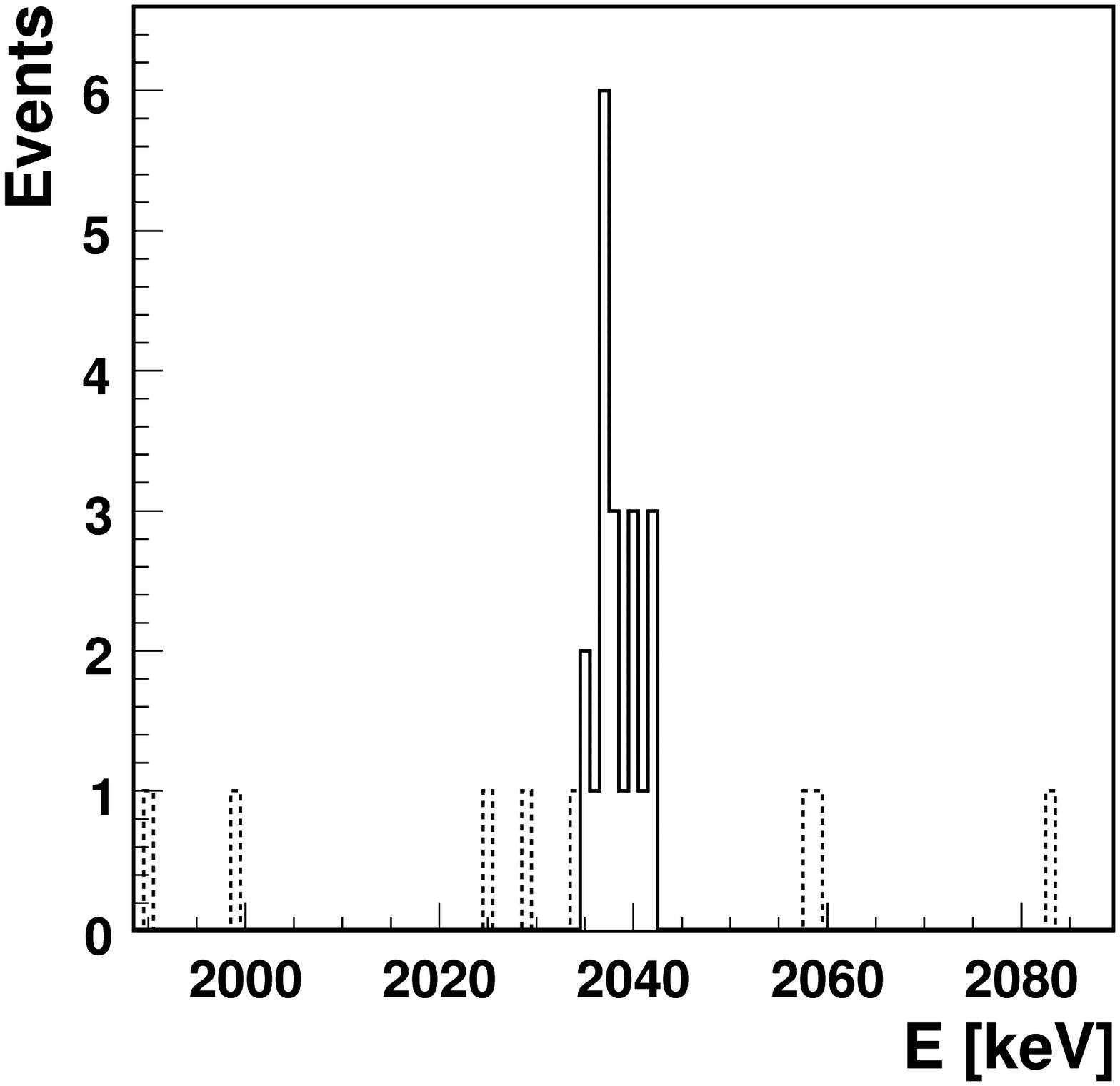,width=\textwidth}}
\end{minipage}
&
\begin{minipage}[ht!]{0.48\textwidth}
\mbox{\epsfig{file=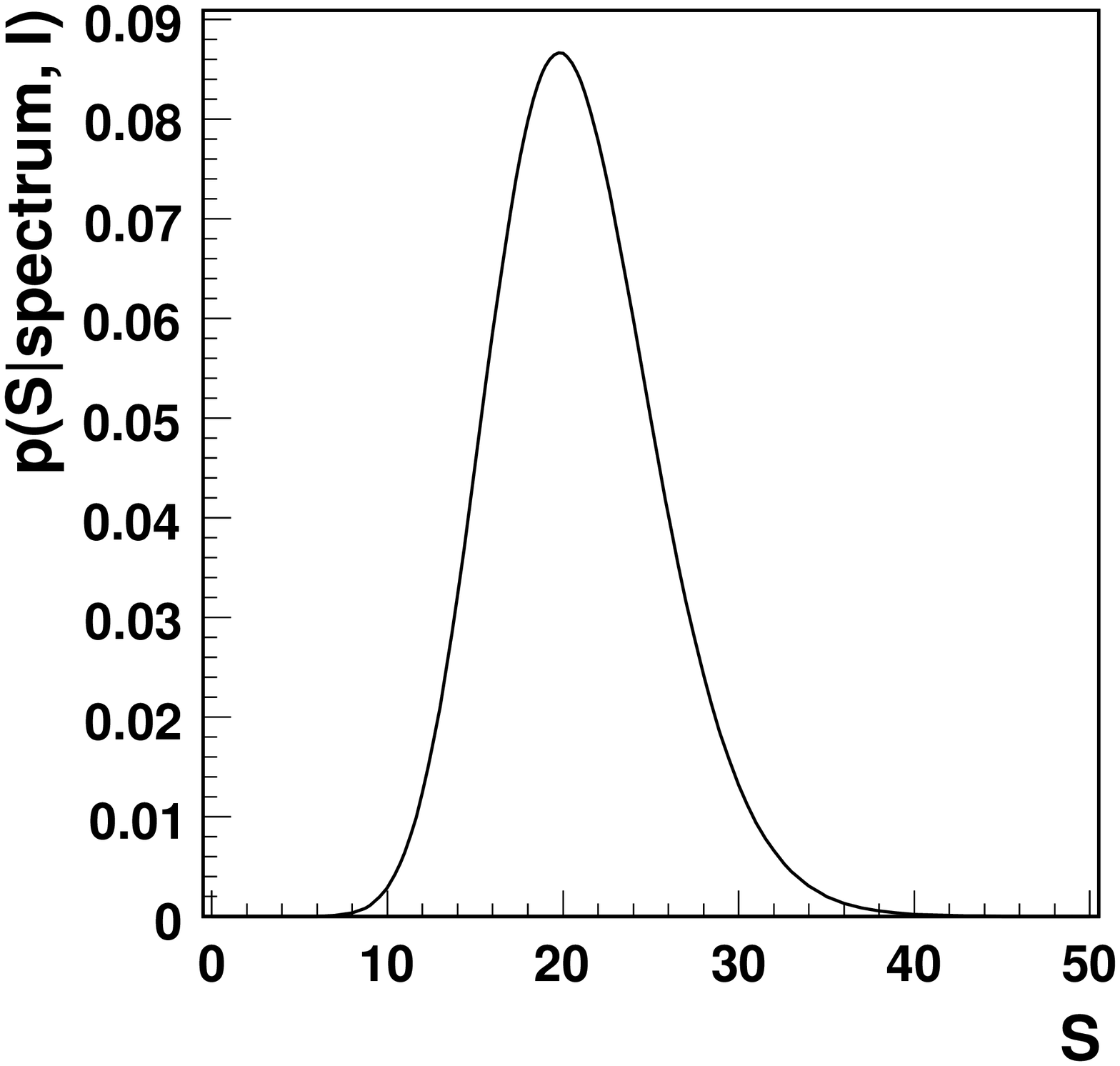,width=\textwidth}}
\end{minipage} \\
\\
\begin{minipage}[ht!]{0.48\textwidth}
\mbox{\epsfig{file=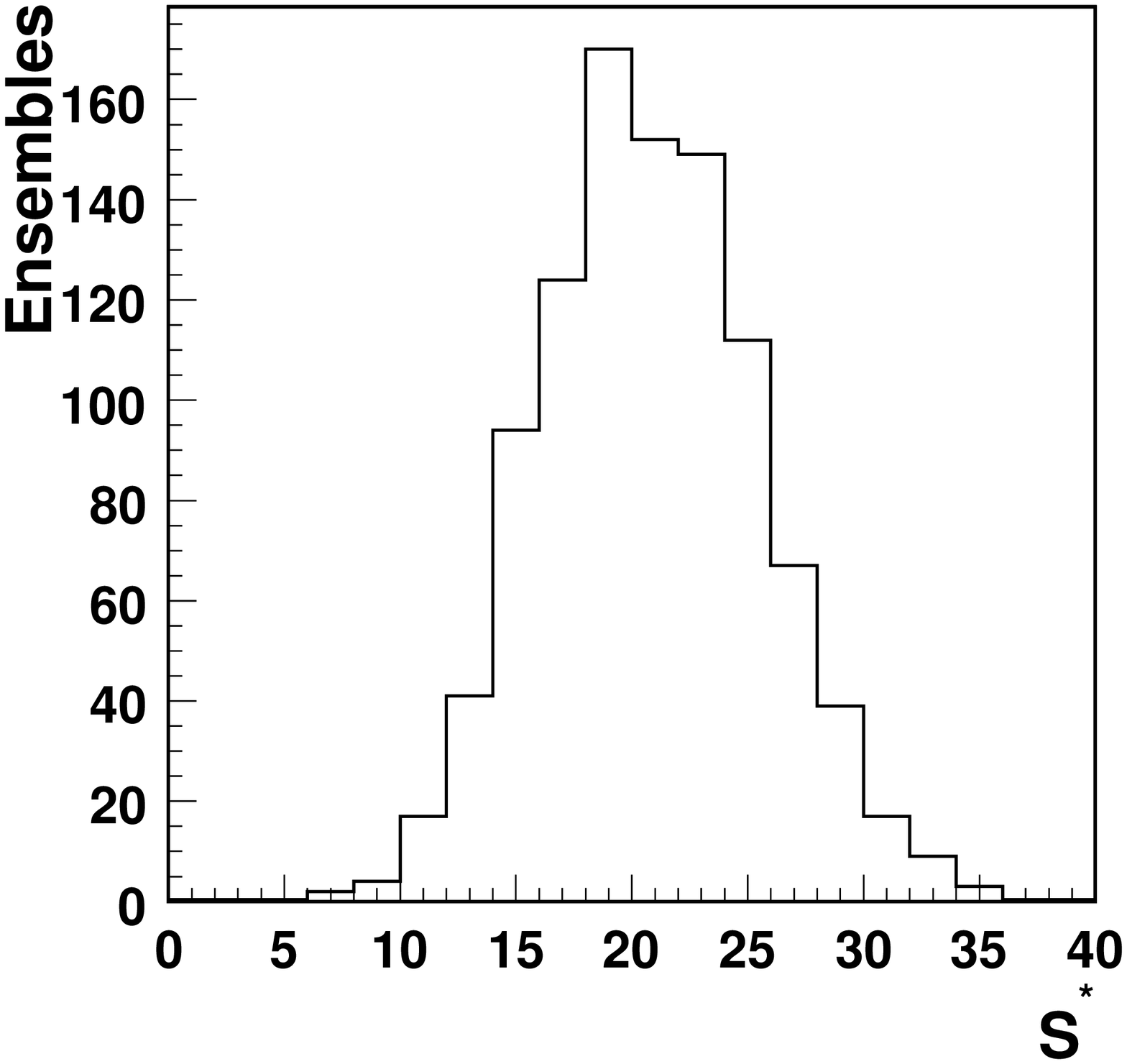,width=\textwidth}}
\end{minipage}
&
\begin{minipage}[ht!]{0.48\textwidth}
\mbox{\epsfig{file=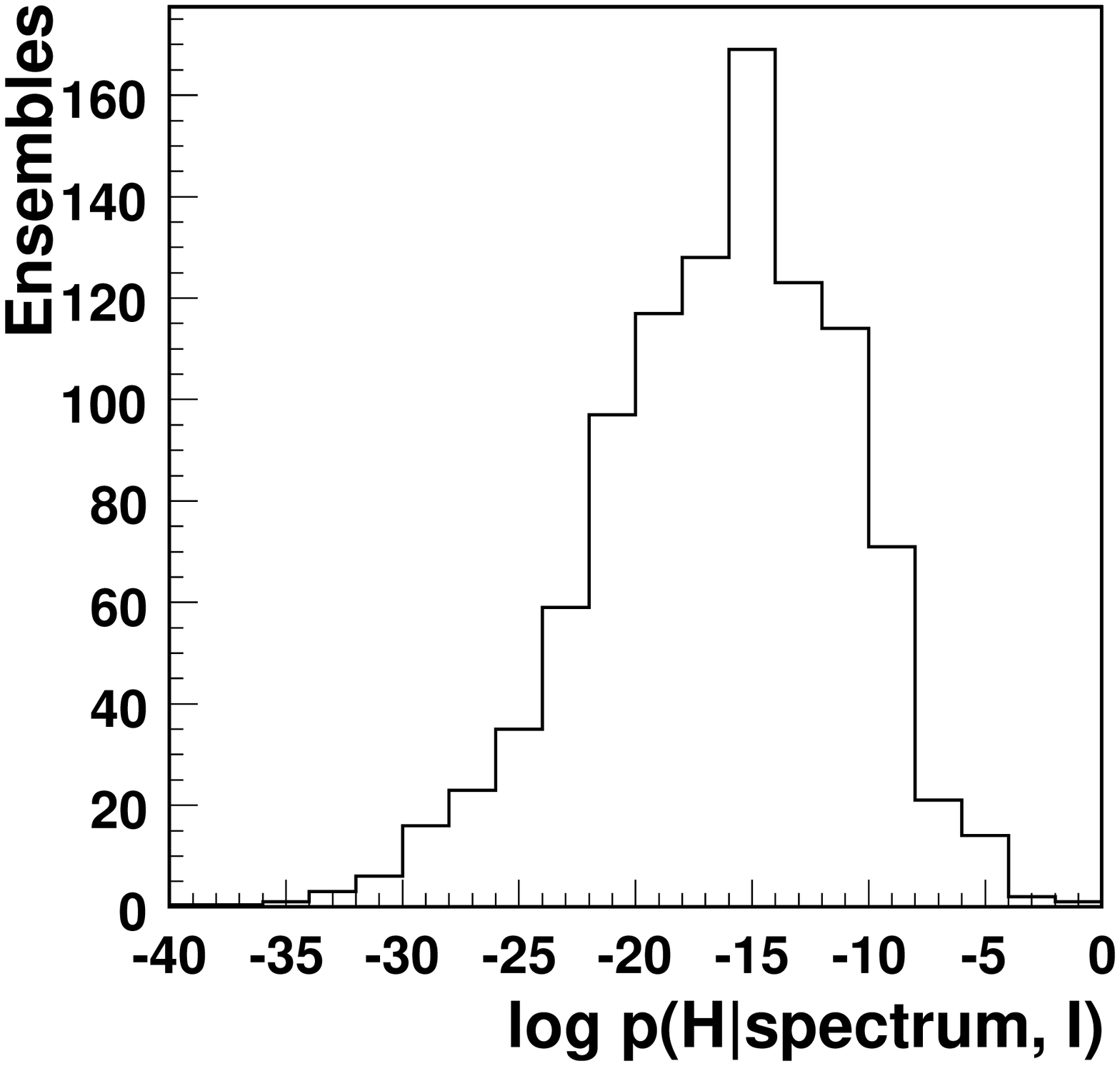,width=\textwidth}}
\end{minipage} 
\end{tabular}
\caption{The spectrum (top, left) was randomly generated under the 
assumptions of a half-life of $2\cdot10^{25}$~years, a background
index of $10^{-3}$~counts/(kg$\cdot$keV$\cdot$y) and an exposure of
100~kg$\cdot$years. The signal events are indicated by a solid line,
the background events by a dashed line. The probability density for
$S$ (top, right) peaks at $19.8$ which is consistent with the number
of signal events, 20, in the spectrum. The distribution of the
estimated number of signal events (bottom, left) as well as the
distribution of the $\log p(H|\textrm{spectrum})$ (bottom, right) are
calculated from ensembles generated under the same assumptions.
\label{fig:example_signal}}
\end{figure}

As an example, Figure~\ref{fig:example_signal} (top, left) shows a
spectrum from Monte Carlo data generated under the assumptions of a
half-life of $2\cdot10^{25}$~years, a background index of
$10^{-3}$~counts/(kg$\cdot$keV$\cdot$y) and an exposure of
100~kg$\cdot$years. This corresponds to $S_{0} = 20.5$ and $B_{0} =
10.0$. The (20) signal and (8) background events are indicated by a
solid and dashed line, respectively.  Figure~\ref{fig:example_signal}
(top, right) shows $p(S|\textrm{spectrum})$ for the same spectrum. The
mode of the distribution is $S^*=19.8$, consistent with the number of
signal events in the spectrum.  Figure~\ref{fig:example_signal}
(bottom, left) shows the distribution of $S^*$ for 1000~ensembles
generated under the same assumptions. The average number of $S^*=
20.3$, in agreement with the average number of generated signal
events, $20.4$.  Figure~\ref{fig:example_signal} (bottom, right) shows
the distribution of the $\log p(H|\textrm{spectrum})$ for ensembles
generated under the same assumptions. More than 97\% of the ensembles
have a probability $p(H|\textrm{spectrum})$ of less than 0.01\%. I.e.,
a discovery could not be claimed for less than 3\% of experiments
under these conditions.

\begin{figure}[ht!]
\begin{tabular}{cc}
\begin{minipage}[ht!]{0.48\textwidth}
\mbox{\epsfig{file=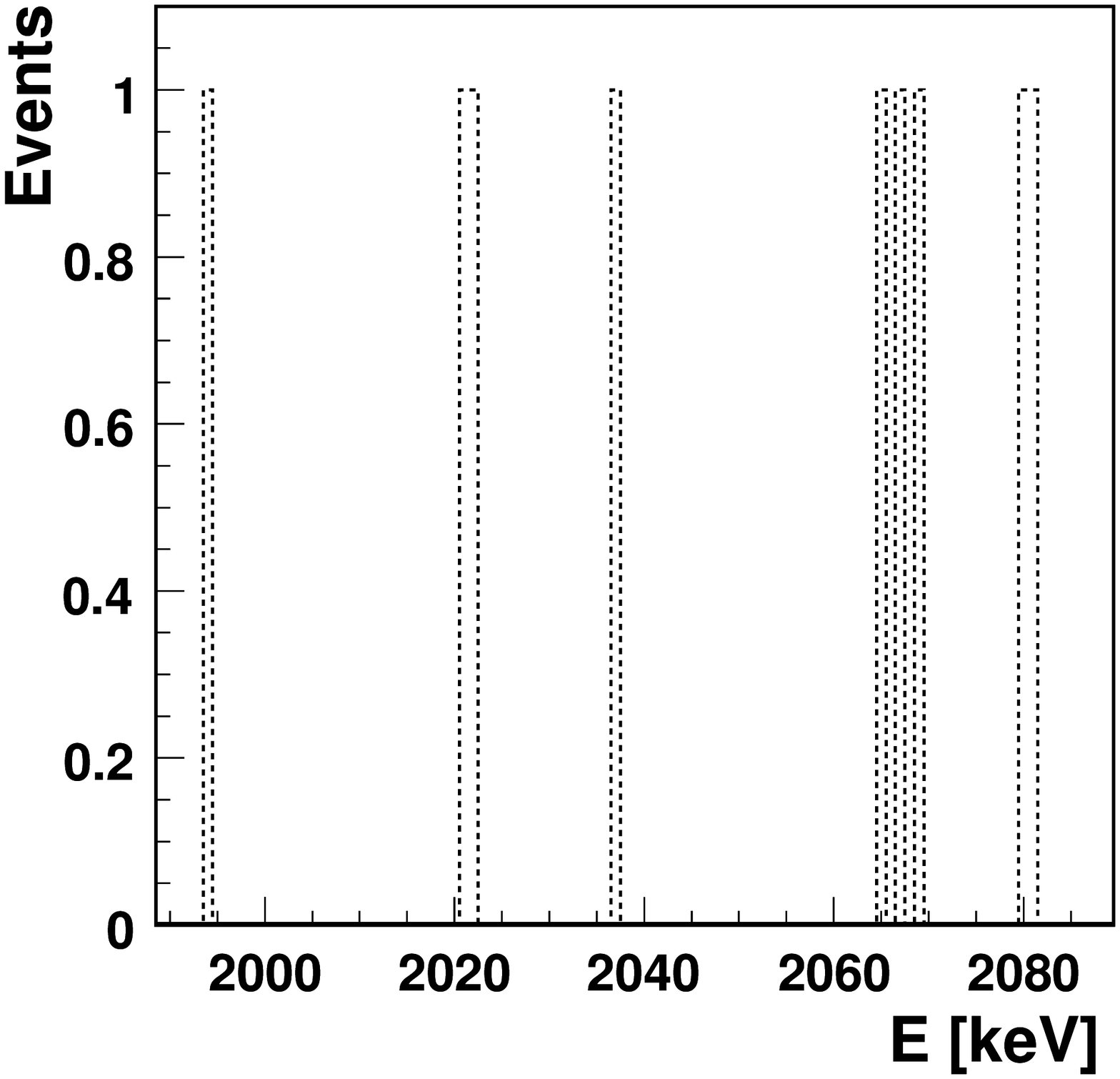,width=\textwidth}}
\end{minipage}
&
\begin{minipage}[ht!]{0.48\textwidth}
\mbox{\epsfig{file=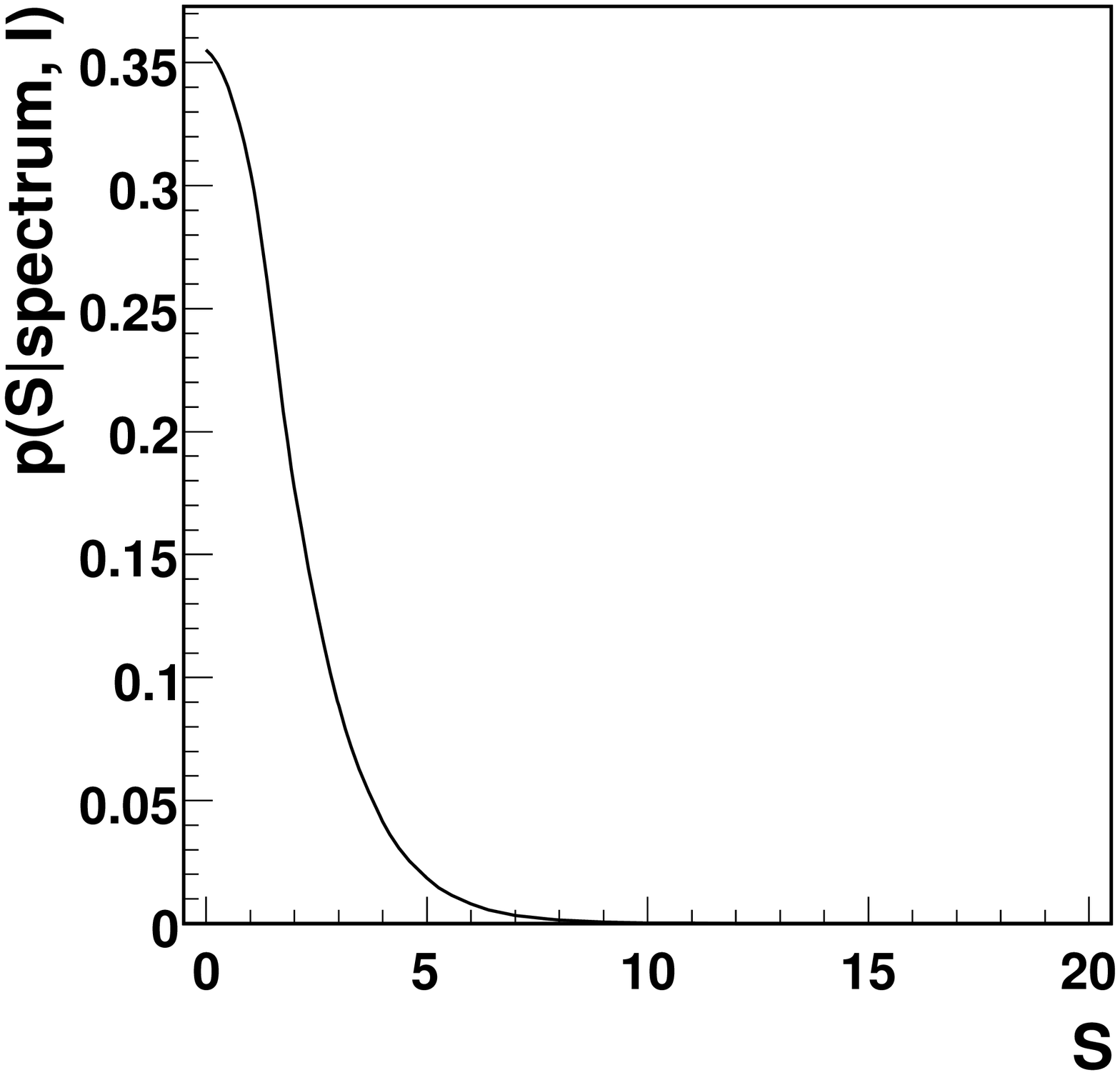,width=\textwidth}}
\end{minipage} \\
\\
\begin{minipage}[ht!]{0.48\textwidth}
\mbox{\epsfig{file=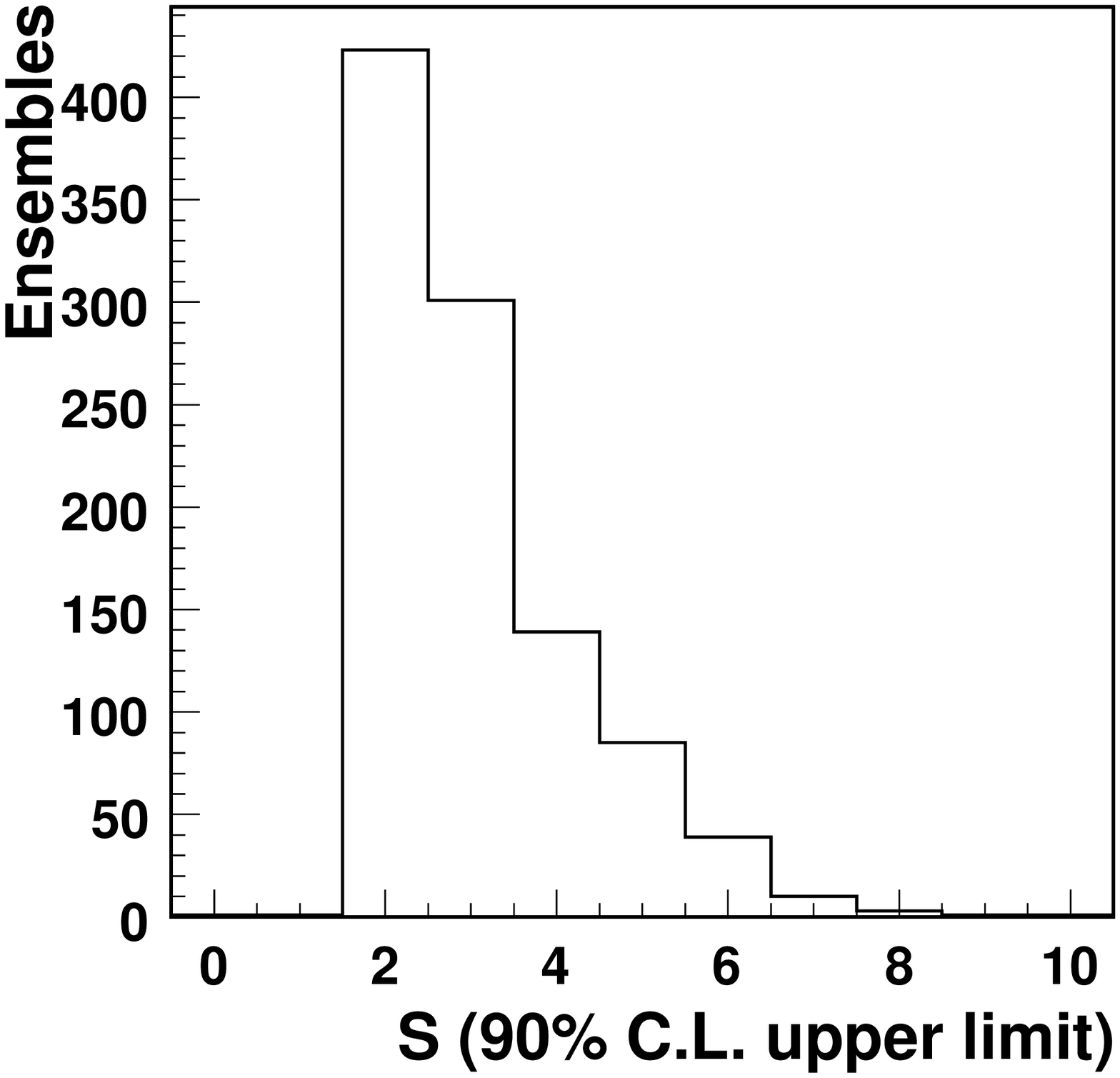,width=\textwidth}}
\end{minipage}
&
\begin{minipage}[ht!]{0.48\textwidth}
\mbox{\epsfig{file=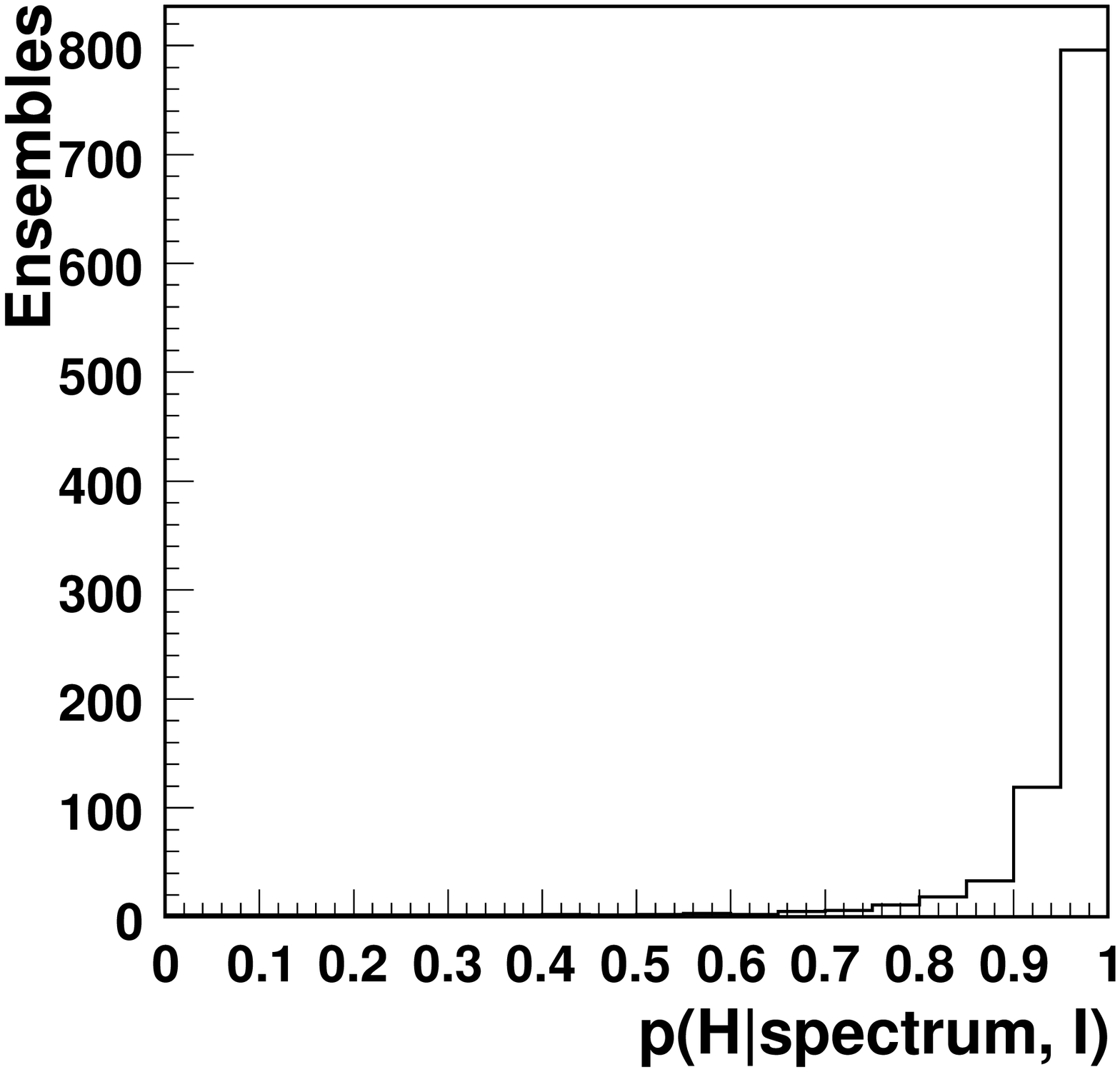,width=\textwidth}}
\end{minipage} 
\end{tabular}
\caption{The spectrum (top, left) was randomly generated under the 
assumptions of a background index of
$10^{-3}$~counts/(kg$\cdot$keV$\cdot$y) and an exposure of
100~kg$\cdot$years. No signal events are present in the spectrum. The
probability density for $S$ for the same spectrum (top, right) peaks
at 0. The distribution of the limit (90\% probability) of the signal
contribution (bottom, left) as well as the distribution of the
$p(H|\textrm{spectrum})$ (bottom, right) are calculated from ensembles
generated under the same assumptions.
\label{fig:example_background}}
\end{figure}

In order to simulate the case in which only lower limits on the
half-life of the $0\nu\beta\beta$-process are set, ensembles are
generated without signal contribution, i.e. $S_{0}=0$. As an example,
Fig.~\ref{fig:example_background} (top, left) shows a spectrum from
Monte Carlo data generated under the assumptions of a background index
of $10^{-3}$~counts/(kg$\cdot$keV$\cdot$y) and an exposure of
100~kg$\cdot$years. No signal events are present in the spectrum.

Figure~\ref{fig:example_background} (top, right) shows the
marginalized probability density for $S$, $p(S|\textrm{spectrum})$,
for the same spectrum. The mode of $S$ is 0 events.

Figure~\ref{fig:example_background} (bottom, left) shows the
distribution of the limit (90\% probability) of the signal
contribution for 1000~ensembles generated under the same
assumptions. The average limit is 3.1.

Figure~\ref{fig:example_background} (bottom, right) shows the
distribution of the $p(H|\textrm{spectrum})$ for ensembles generated
under the same assumptions. For none of the ensembles could a
discovery be claimed.
 
\clearpage 

\subsection{Sensitivity} 
 
\begin{figure}[ht!]
\center
\mbox{\epsfig{file=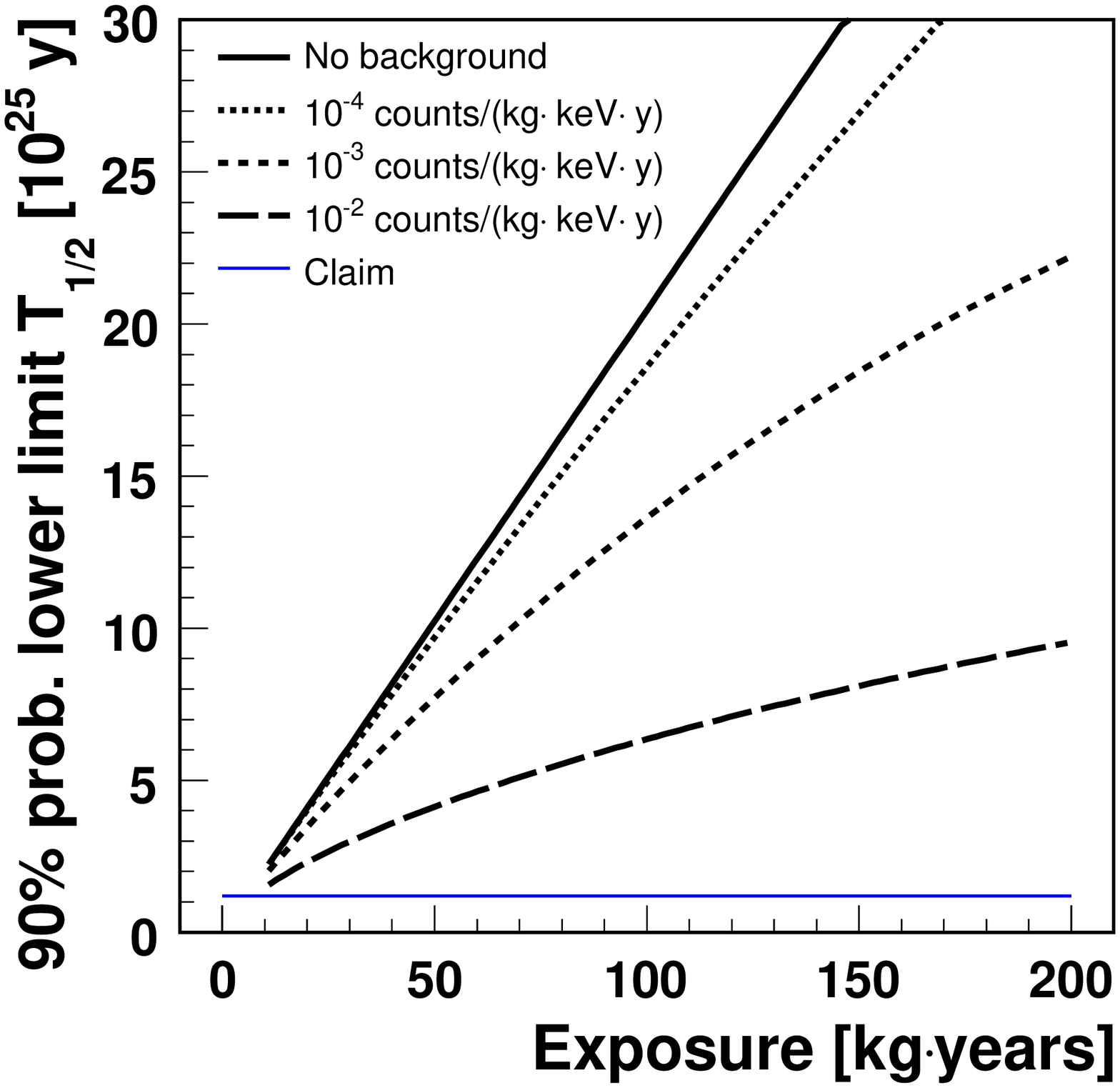,width=0.7\textwidth}} \\ 
\mbox{\epsfig{file=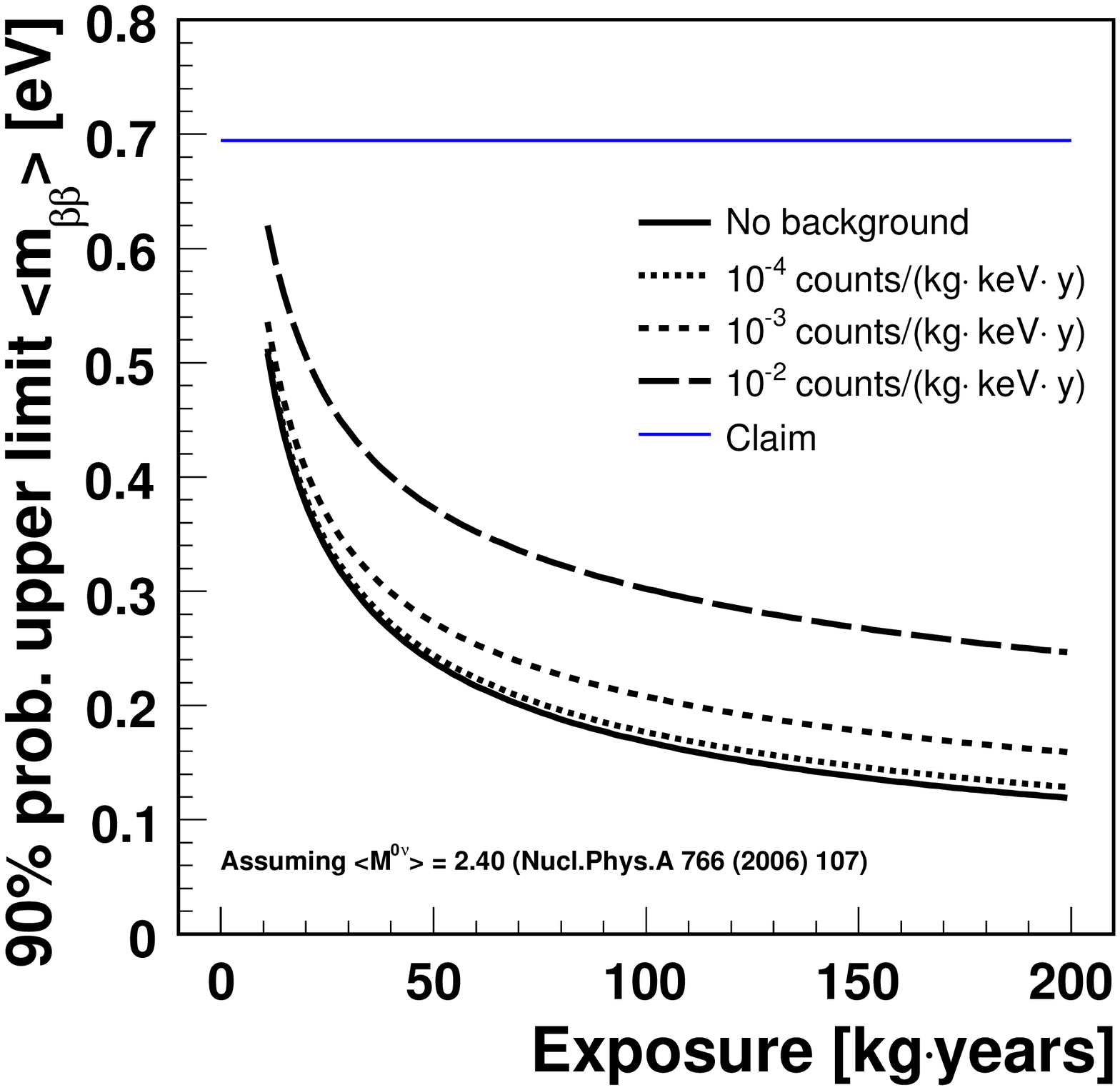,width=0.7\textwidth}}
\caption{The upper plot shows the expected 90\% probability lower limit on the 
half-life for neutrinoless double beta decay versus the exposure under
different background conditions. Also shown is the half-life for the
claimed observation~\cite{Klapdor}. The lower plot shows the expected
90\% probability upper limit on the effective Majorana neutrino mass
versus the exposure under different background conditions. The
effective Majorana neutrino mass for the claimed observation is also
shown. All mass values were determined from the half-life using the
matrix element reported in~\cite{matrix}.
\label{fig:limit}}
\end{figure}

For the ensembles generated without signal contribution the mean of
the 90\% probability lower limit on the half-life is shown in
Figure~\ref{fig:limit} as a function of the exposure for different
background indices. In case no background is present the limit scales
linearly with the exposure. With increasing background contribution
the limit on the half-life increases more slowly. For the envisioned
background index of $10^{-3}$~counts/(kg$\cdot$keV$\cdot$y) and an
expected exposure of 100~kg$\cdot$years an average lower limit of
$T_{1/2}>13.5\cdot10^{25}$~years can be set. For the same exposure,
the average lower limit is $T_{1/2}>6.0\cdot10^{25}$~years and
$T_{1/2}>18.5\cdot10^{25}$~years for background indices of
$10^{-2}$~counts/(kg$\cdot$keV$\cdot$y) and
$10^{-4}$~counts/(kg$\cdot$keV$\cdot$y), respectively.

Using the nuclear matrix elements quoted in~\cite{matrix} the lower
limit on the half-life of the $0\nu\beta\beta$-process can be
translated into an upper limit on the effective Majorana neutrino
mass, $\langle m_{\beta\beta} \rangle$, via
\begin{equation} 
\langle m_{\beta\beta} \rangle = (T_{1/2} \cdot G^{0\nu})^{-1/2} \cdot \frac{1}{\langle M^{0\nu}\rangle}, 
\label{eqn:mass}
\end{equation} 

where $G^{0\nu}$ is a phase space factor and $\langle M^{0\nu}\rangle$
is the nuclear matrix element. Figure~\ref{fig:limit} also shows the
expected 90\% probability upper limit on the effective Majorana
neutrino mass as a function of the exposure. With a background index
of $10^{-3}$~counts/(kg$\cdot$keV$\cdot$y) and an exposure of
100~kg$\cdot$years, an upper limit of $\langle m_{\beta\beta} \rangle
<200$~meV could be set assuming no $0\nu\beta\beta$-events are
observed. \\

\begin{figure}[ht!]
\center
\mbox{\epsfig{file=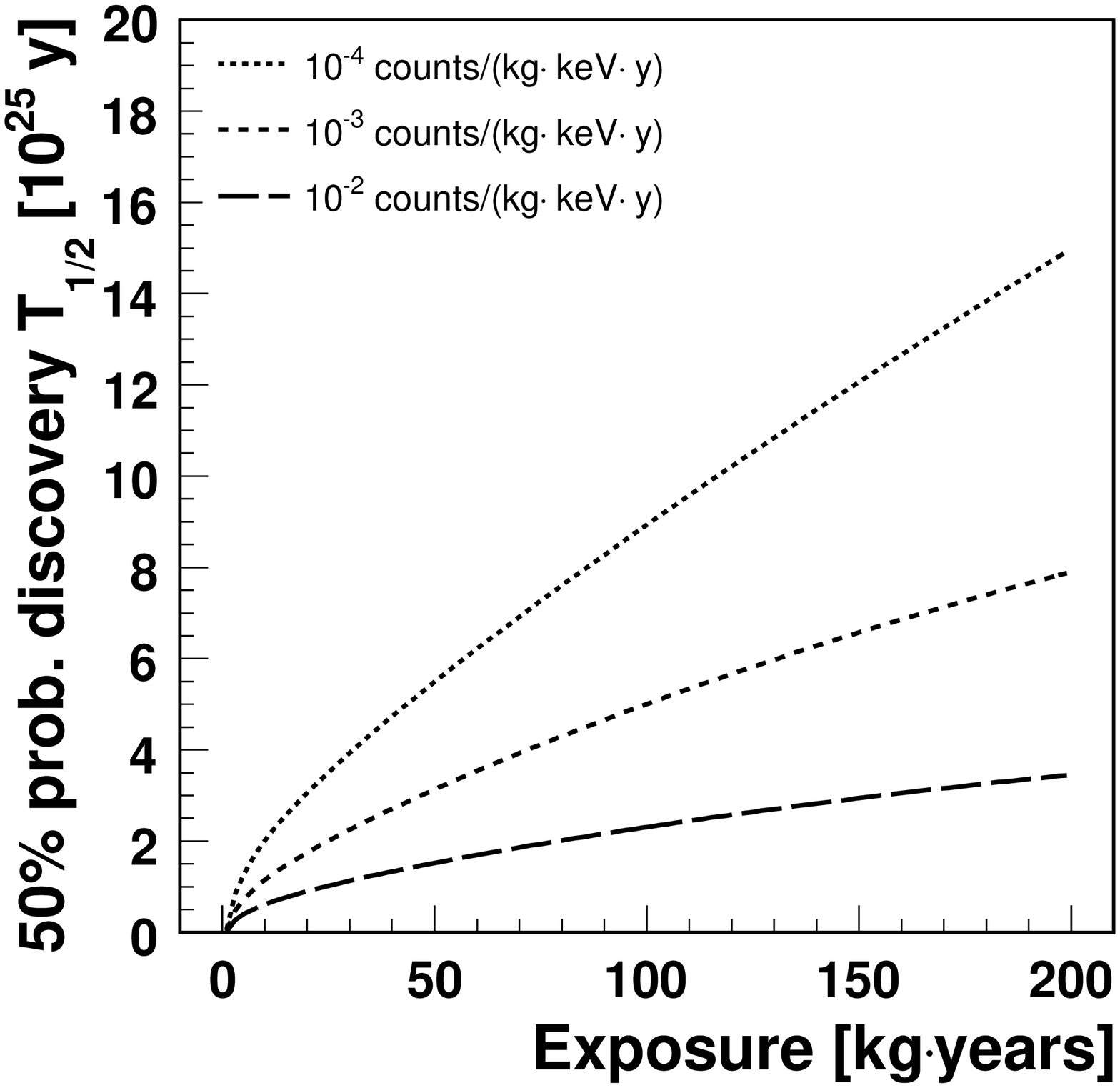,width=0.7\textwidth}}
\mbox{\epsfig{file=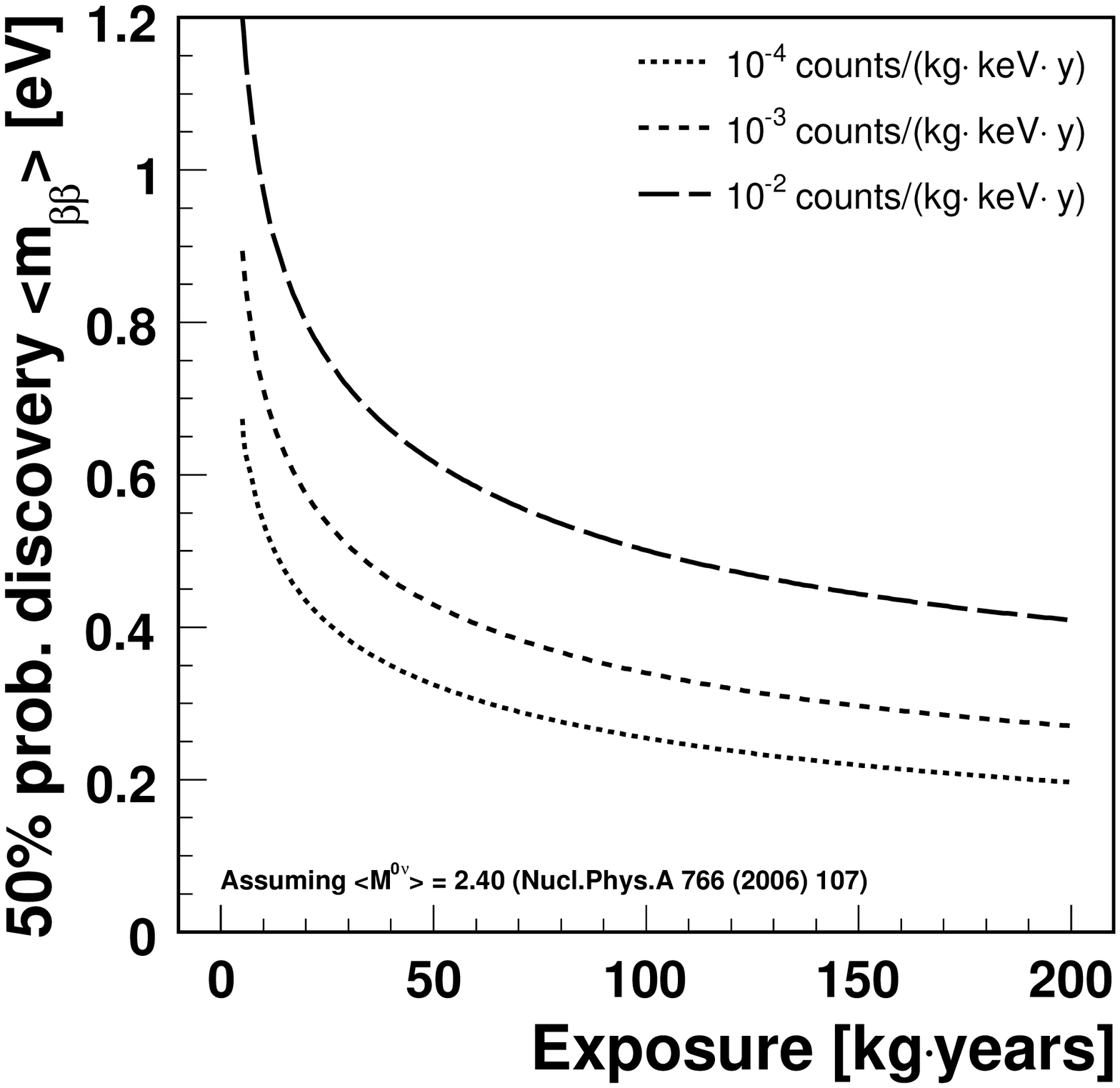,width=0.7\textwidth}}
\caption{Top: the half-life for which 50\% of the experiments would 
report a discovery, i.e. a probability that the spectrum is due to
background processes only, $p(H|\textrm{spectrum})$, of less than
0.01\%, is plotted versus the exposure under different background
conditions. Bottom: the effective Majorana neutrino mass for which
50\% of the experiments would report a discovery versus the exposure
under different background conditions. The mass was determined from
the half-life using the matrix element reported in~\cite{matrix}.
\label{fig:discovery}}
\end{figure}

Figure~\ref{fig:discovery} shows the half-life for which 50\% of the
experiments would report a discovery of neutrinoless double beta decay
as a function of the exposure for different background indices. For
the envisioned background index of
$10^{-3}$~counts/(kg$\cdot$keV$\cdot$y) and an expected exposure of
100~kg$\cdot$years this half-life is $5\cdot10^{25}$~years. \\
 
Using the same matrix elements from reference~\cite{matrix}, the
half-life is transformed into an effective Majorana neutrino mass. The
mass for which 50\% of the experiments would report a discovery is
shown in Figure~\ref{fig:discovery} (bottom) as a function of the
exposure and for different background conditions. For an exposure of
100~kg$\cdot$years and a background index of
$10^{-3}$~counts/(kg$\cdot$keV$\cdot$y) neutrinoless double beta decay
could be discovered for an effective Majorana neutrino mass of 350~meV
(with a 50\% probability).
 
\clearpage 

\subsection{Influence of the prior probabilities} 

In order to study the influence of the prior probabilities on the
outcome of the experiment, the prior probability for the number of
expected signal events, $p_{0}(S)$, was varied. Three different prior
probabilities were studied:

\begin{itemize} 
\item flat prior: $p_{0}(S) \propto const.$~,
\item pessimistic prior: $p_{0}(S) \propto e^{-S/10}$~,
\item peaking prior: $p_{0}(S) \propto e^{1 - \tilde{S}/S} / S^{2}$, 
\end{itemize} 

where $\tilde{S}$ is the number of events corresponding to a half-life
of $1.2\cdot10^{25}$~years and $S<S_{\mathrm{max}}$. For a background
index of $10^{-3}$~counts/(kg~keV~y) and an exposure of 100~kg~years
the limit strongly depends on the chosen prior. For the pessimistic
prior probability the limit which can be set on the half-life is about
10\% higher than that for the flat prior probability. In comparison,
the peaking prior gives a 50\% lower limit compared to the flat
prior. This study makes the role of priors clear.  If an opinion is
initially strongly held, then substantial data is needed to change it.
In the scientific context, consensus priors should be strived for. \\


\section{Conclusions}
\label{section:conclusions}
 
An analysis method, based on Bayes' Theorem, was developed which can
be used to evaluate the probability that a spectrum can be explained
by background processes alone, and thereby determine whether a signal
process is necessary.  A criterion for claiming evidence for, or
discovery of, a signal was proposed. Monte Carlo techniques were
described to make predictions about the possible outcomes of the
experiments and to evaluate the sensitivity for the process under
study. \\
 
As an example the method was applied to the case of the {\sc GERDA}
experiment for which the sensitivity to neutrinoless double beta decay
of $^{76}$Ge was calculated. With a background index of
$10^{-3}$~counts/(kg$\cdot$keV$\cdot$y) and an exposure of
100~kg$\cdot$years the sensitivity of the half-life of the
$0\nu\beta\beta$-process is expected to be
$13.5\cdot10^{25}$~years. \\


\addcontentsline{toc}{section}{Bibliography}
%


\end{document}